\documentclass[aps,prb,showpacs,twocolumn]{revtex4}

\usepackage{graphicx}
\begin{document}

\title
{Boundary Energies and the Geometry of Phase Separation in Double--Exchange
Magnets}
\author{D. I. Golosov}
\email{golosov@phys.huji.ac.il}
\affiliation{Racah Institute of Physics,
The Hebrew University, Jerusalem 91904, Israel.}
\author {D. Orgad}
\affiliation{Racah Institute of Physics,
The Hebrew University, Jerusalem 91904, Israel.}

\date{\today}

\begin{abstract}
We calculate the energy of a boundary between ferro- and antiferromagnetic
regions in a phase separated double-exchange magnet in two and three
dimensions.
The orientation dependence of this energy can
significantly affect the geometry
of the phase-separated state in two dimensions, changing the droplet shape
and possibly stabilizing a striped arrangement within a certain range of
the model parameters. A similar effect, albeit weaker, is also
present in three dimensions. As a result,
a phase-separated system near the percolation threshold is expected to possess
intrinsic hysteretic transport properties, relevant in the context of recent
experimental findings.
\typeout{polish abstract}
\end{abstract}

\maketitle

Phase separation is ubiquitous in systems involving multiple types of degrees
of freedom
and opposing interactions.
Among the growing number of compounds which are known to exhibit
experimental signatures consistent with this phenomenon one finds
the doped manganese oxides \cite{Nagaevbook,Dagottobook} [renowned due to
their colossal magnetoresistance (CMR)], as well as lightly-doped magnetic
semiconductors \cite{Nagaevbook} ({\it e.g.}, Eu-based).
The basic physics of both of these systems
involves conduction electrons that hop between sites in a crystal,
whose ionic (localized) core spins are coupled to the electronic spins
via a strong on-site ferromagnetic Hund's exchange interaction.
Under broad conditions, the energy of the Fermi sea of conduction electrons
is then minimized when the core spins are ordered ferromagnetically,
giving rise to {\it double-exchange} ferromagnetism\cite{deGennes}.
This ordering is opposed by strong antiferromagnetic tendencies present
in such systems, which originate from Heisenberg superexchange
between the core spins, and which may also be complemented by other,
band-structure and orbital-related, mechanisms.
Phase separation resolves this competition by dividing the
system into regions of different magnetic orders and electronic band-widths.
We will be concerned with the case where the
ferromagnetic regions in which the electrons are free to move
(and thereby to lower their kinetic energy)
co-exist with electron-poor domains
where antiferromagnetism prevails.

The present article focuses on  the properties
of the boundaries between different regions in the sample.
It is shown that with increasing strength
of antiferromagnetism the boundary energy becomes increasingly dependent
on the orientation of the boundary with respect to the lattice.
Thus, when the Eu chalcogenides (which have N\'{e}el temperatures
of the order of 10 K or below) are doped with carriers, the latter are
predicted to form droplets of circular or spherical shape.
In the CMR manganates, on the other hand, antiferromagnetism is much more
pronounced and we expect a strong dependence of the boundary energy on
orientation.
This fact, in turn, is likely to contribute to the unusual
transport properties of the CMR manganates near percolation threshold
including hysteresis, memory and switching effects \cite{bulk,films}.

Although to the best of our knowledge such considerations have not been
presented before, it appears reasonable to expect that at the qualitative
level these are not specific for the simplified model considered below,
nor indeed for double-exchange magnetism. Support for this view can be
drawn from the numerical studies of phase separation in the $t-J$ and Hubbard
models, showing diamond-shaped islands \cite{Berkovits} or an anisotropy
in the stripe direction \cite{Hubbard}. Rigorous results on the preferred
wall direction between ferromagnetic domains in the Falicov--Kimball
model \cite{Nachtergaele} are likely to be indicative also of the behavior
of phase separation boundaries in this model.

We begin with the standard double-exchange Hamiltonian,
\begin{eqnarray}
{\cal H}=&&-\frac{t}{2} \sum_{\langle i,j \rangle,\alpha}
\left(c^\dagger_{i \alpha}c_{j \alpha} +c^\dagger_{j \alpha}c_{i
\alpha}\right) \nonumber \\
 &&-\frac{J_H}{2S} \sum_{i, \alpha, \beta} \vec{S}_i
\vec{\sigma}^{\alpha \beta} c^\dagger_{i\alpha} c_{i\beta}
+\frac{J}{S^2}\sum_{\langle i, j \rangle} \vec{S}_i \vec{S}_j\,.
\label{eq:Ham}
\end{eqnarray}
Here  $c_{j \alpha}$ ($\alpha=\uparrow, \downarrow$) are the electron
annihilation operators,
$\vec{S}_i$ are the operators of the core spins
located at the sites of a square (or simple cubic) lattice, and
the vector $\vec{\sigma}^{\alpha \beta}$ is composed of Pauli
matrices. The ionic spins, which are treated classically (assuming $S \gg 1$),
interact with each other via a nearest-neighbor superexchange $J$ and with
the spins of the conduction electrons via a large Hund's coupling
$J_H$. 
We consider the limit\cite{limitcom} $J_H\rightarrow \infty$ and omit all
peculiarities of lattice and orbital structure (which vary between
compounds). While the long-range Coulomb interaction is absent from the
Hamiltonian its effects will be discussed qualitatively.
Throughout this paper we assume zero temperature
and take the lattice spacing, the bare hopping amplitude $t$, and $\hbar$
to be 1.
Since the resulting model is symmetric with
respect to quarter-filling $x=0.5$, we treat the $x<0.5$ case only.

As $J_H\rightarrow \infty$, there is only one spin
state available for a conduction electron at each lattice site. The carriers
can therefore be treated as spinless, and their nearest-neighbor hopping
amplitude depends on the local spin configuration via the double-exchange
mechanism\cite{Anderson}. In particular, it equals 1
when the ionic spins at the two neighboring sites are parallel to each other,
and it vanishes in the case of anti-parallel spins.
When a certain number of carriers, $N_e= xN$, where $N$ is the number of
sites in the system, are doped into an
insulating (antiferromagnetic) double-exchange magnet, it proves energetically
favorable for all these carriers to concentrate in one part of the sample.
The magnetic ordering in this part is changed to a ferromagnetic one, while
the rest of the sample remains antiferromagnetic and free of conduction
electrons (as long as $x$ and $J$ are not too large, see below).
The optimal value of the carrier concentration in the ferromagnetic region,
$x_{FM}>x$, is determined by the condition
\begin{equation}
\Omega_{AFM}=\Omega_{FM}\,\,,
\label{eq:equilibr}
\end{equation}
where in the case of dimensionality $D$ the
thermodynamic potentials of the antiferromagnetic and ferromagnetic
phases are given by $\Omega_{AFM}(J)=-DJ$ and
$\Omega_{FM}(J,\mu)= \int_{-D}^\mu (\epsilon-\mu) \nu_D(\epsilon) d \epsilon
+DJ$.
Here the chemical potential is denoted by $\mu - J_H/2$ and $\nu_D(\epsilon)$
is the density of states for the 2D (3D) tight-binding electronic dispersion,
$\epsilon_{\vec{k}}=-\cos k_x -\cos k_y (-\cos k_z)$.

The values of $\mu$ and
$x_{FM}=\int_{-D}^\mu
\nu_D(\epsilon) d \epsilon$ depend solely on $J$
(see the dotted lines in Fig. \ref{fig:tensions} below). When the value of
the nominal density $x$ reaches
$x_{FM}(J)$, the entire sample turns ferromagnetic. 
However, an investigation of the energy balance and stability of different
phases shows that thermodynamic equilibrium between N\'{e}el and ferromagnetic
phases is possible only as long as $x_{FM}$ is not too large, $x_{FM}<x_c$ or
equivalently $J<J_c$. While the precise value of $J_c$ is not known, a
variational study\cite{jap02} in 2D  shows that $x_c < 0.245$,
corresponding
to $J_c < 0.036$. In 3D, $x_c<0.291$ or $J_c<0.035$,
at which
point the spin stiffness in the ferromagnetic phase turns negative.
If the value of $J$ exceeds $J_c$ the magnetic
ordering in either the electron-rich or the electron-poor regions of the
sample will differ from that of a
ferromagnet or a N\'{e}el antiferromagnet.
Rather, one of the many
possible antiferromagnetic phases characterized by non-vanishing conduction
electron density may be stabilized (see Ref. \onlinecite{jap02}
for examples).
Throughout the present article, we assume that $J$ is smaller than
$J_c$.

\begin{figure}[h!!!]
\includegraphics[angle=0,width=3.2in]{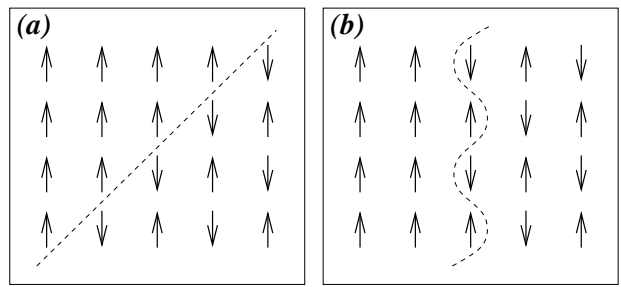}
\caption{Diagonal {\it (a)} and vertical {\it (b)} boundaries in the 2D case.}
\label{fig:walls}
\end{figure}

The presence of phase separation inevitably raises questions regarding the
structure of the boundary between the phases.
In an earlier
article\cite{jap02}, it was suggested that the most energetically
favorable structure of such a boundary corresponds to an abrupt
change in the magnetic order. For the case of
phase separation into ferromagnetic and  N\'{e}el phases, which is
of interest to us here, straight boundaries running in the
direction of a lattice diagonal in a 2D system
were considered [see Fig. \ref{fig:walls} {\it (a)}].
The energy of such a boundary per unit length is given by\cite{prb03}
\begin{eqnarray}
\sqrt{2}W_d^{(2D)}(J)\!&=&\!\frac{\sqrt{4-\mu^2}}{2 \pi}-
\frac{|\mu|}{2 \pi}{\rm arccos}
\frac{|\mu|}{2} -\mu x_{FM}\label{eq:2Ddiag}\\
\!&-&\!\frac{4}{\pi^2} {\cal E}\left(\sqrt{1-\frac{\mu^2}{4}}\right)
+\frac{\mu^2}{\pi^2}{\cal K}\left(\sqrt{1-\frac{\mu^2}{4}}\right),
\nonumber
\end{eqnarray}
where ${\cal K}$ and ${\cal E}$ are the elliptic integrals \cite{numericscom}.

In 3D, the analogue of such an abrupt diagonal boundary is an abrupt planar
boundary perpendicular to a main diagonal [direction
$(1,1,1)$ or equivalents] of the cubic lattice. The calculations of Ref.
\onlinecite{prb03} can be generalized to the 3D case, yielding the boundary
energy,
\begin{eqnarray}
\nonumber
&&\!\!\!W_d^{(3D)}= \frac{1}{8} \int [\mu + \lambda(\vec{k}_\bot)]
\theta \left(\mu+
\lambda(\vec{k}_\bot)\right)\! \frac{d^2 k_\bot}{4 \pi^2}- 2\sqrt{3} J,
\label{eq:3Ddiag} \\
&&\!\!\!\lambda^2(\vec{k}_\bot) =3+2 \cos k_1+4\cos \frac{\sqrt{3}\, k_2}{2}
\cos \frac{k_1}{2} ,
\end{eqnarray}
per unit area. In Eq. (\ref{eq:3Ddiag}), the integration variable
$\vec{k}_\bot=\{k_1,k_2\}$  varies over the first Brillouin zone of a
triangular lattice of unit spacing. The $k_1$ axis is chosen in one of
the nearest-neighbor directions.

In the case of small $J \ll 1$, these somewhat cumbersome expressions reduce to
\begin{equation}
W_d\approx\left\{ \begin{array}{ll}
\frac{4\sqrt{2}}{3\pi^{1/4}}
J^{3/4}\!-2 \sqrt{2}J & {\rm in \,\, 2D,}\\ \,&\, \\
\!\frac{3^{8/5}5^{4/5}}
{16\cdot 2^{2/5}}
\pi^{3/5}J^{4/5}
\!-2 \sqrt{3}J & {\rm in \,\,3D.}
\end{array} \right.
\label{eq:smallJdiag}
\end{equation}

In both 2D (Refs. \onlinecite{jap02,numericscom}) and 3D cases,
one can verify that
tilting the spins adjacent to a diagonal boundary (effectively spreading
it over two or three lattice links) increases the boundary energy. This
strongly supports the notion of an abrupt inter-phase boundary. The fact
that the boundary is sharp (as opposed to a smooth Bloch-like wall) implies
the absence of a continuum limit and hence a non-trivial dependence of the
boundary energy on orientation. We will now explore this property
variationally in more detail.

Specifically, we consider a boundary running perpendicularly to
the $(1,0)$ and $(1,0,0)$ directions in the 2D [Fig.\ref{fig:walls} {\it (b)}]
and 3D cases, respectively.
The energy of such a ``vertical'' wall in $D=2,3$ dimensions can
be evaluated exactly within the same technique (see Appendix). We find
\begin{equation}
W_v= \frac{1}{2}\int_{1-D}^{0}\nu_{D-1}(\epsilon_\bot)d \epsilon_\bot
\int_{-D}^{\mu(J)}\xi(\epsilon,\epsilon_\bot)d\epsilon -DJ.
\label{eq:23Dvert}
\end{equation}
Assuming that $\epsilon<0$,
the spectral shift function $\xi(\epsilon,\epsilon_\bot)$ is given by
\begin{equation}
\xi\!=\!\!\left\{ \begin{array}{ll}
0, & \epsilon_\pm\!<-Q\\
\frac{1}{2} - \frac{2}{\pi}{\rm arc tan}
\frac{\sqrt{Q^2-\epsilon_+^2}}{\sqrt{\epsilon_-^2-Q^2}-2 \epsilon},
& \epsilon_-\!<\!-Q\!<\epsilon_+\!<\!Q\\~&~\\
1 - \frac{2}{\pi}{\rm arc tan}
\frac{\sqrt{Q^2-\epsilon_+^2}+\sqrt{Q^2-\epsilon_-^2}}{-2 \epsilon},
& -Q<\epsilon_\pm<Q\\
1, &\epsilon_-\! <\!-Q,\,\,\epsilon_+>\!Q
\end{array} \right.
\label{eq:xivert}
\end{equation}
where in the present case $Q=1$ and $\epsilon_\pm=\epsilon \mp \epsilon_\bot$.

At small J, one obtains
\begin{equation}
W_v\approx\left\{ \begin{array}{ll}
\frac{4\sqrt{2}}{3\pi^{1/4}}J^{3/4}-(4-\sqrt{2})J
& {\rm in\,\, 2D,}\\ \,&\,\\
\frac{3^{8/5}5^{4/5}}{16\cdot 2^{2/5}}\pi^{3/5}J^{4/5}
\!-\left( 6-\sqrt{6}\right)J
& {\rm in\,\, 3D.}
\end{array} \right.
\label{eq:smallJvert}
\end{equation}
The leading-order terms in Eqs. (\ref{eq:smallJdiag}) and
(\ref{eq:smallJvert})
originate from the energy of a partition inserted into an ideal gas of
electrons.
In the limit of small $J$ (corresponding to small $x_{FM}$ and to a large Fermi
wavelength) this quantity should not depend on the orientation of the partition
with respect to the lattice, as is manifest from Eqs.
(\ref{eq:smallJdiag}) and (\ref{eq:smallJvert}).
This independence of the boundary energy on orientation at small $J$
is expected to hold generally, and not only for the two directions
considered above.
The droplets of a ferromagnetic phase within an
antiferromagnetic sample will then have a circular (spherical) shape,
obtained by minimizing the boundary length or area \cite{large}.
This
changes with increasing $J$ as we discuss below.

\begin{figure}[ht!!!]
\includegraphics[angle=0,width=3.2in]{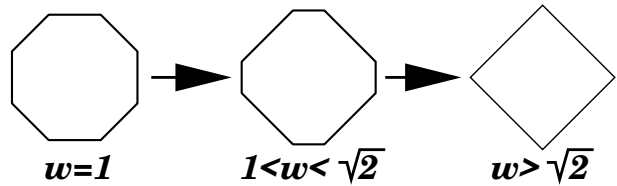}
\caption{Evolution of the droplet shape
in the 2D case.}
\label{fig:circlesquare}
\end{figure}

\noindent {\bf The 2D case}. The energy of a vertical boundary always exceeds
that of the diagonal one [see Fig. \ref{fig:tensions} {\it (a)}], and
with increasing $J$ the ratio  $w=W_v/W_d$ increases.
The droplet shape in 2D then approaches that of
a diamond with rounded corners. The curvature radii at the
corners decrease as $w$ increases. We model this transition
approximately, allowing only for the vertical and diagonal orientations of the
boundary (see Fig. \ref{fig:circlesquare}). Minimizing the boundary energy of
the droplet with respect to the ratio $\alpha$ of the combined length of the
vertical components to that of the diagonal ones while keeping the area of
the droplet constant yields $\alpha=(\sqrt{2}-w)/(\sqrt{2}w-1)$.
As the value of $w$ increases
from $1$ to $\sqrt{2}$ (the latter corresponds to $J \approx 0.030$ or
$x_{FM} \approx 0.221$), the droplet
shape changes from that of a regular
octagon\cite{percent} to that of a diamond.

\begin{figure}[ht!!!]
\includegraphics[angle=0,width=3.41in]{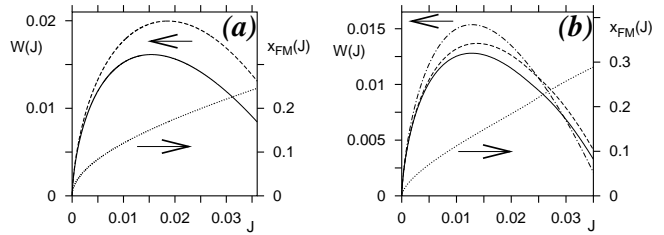}
\caption{The dependence of the diagonal, vertical, and [in {\it (b)}] \{110\}
boundary energies (solid, dashed, and dashed-dotted lines, respectively)
on the value of the superexchange coupling $J$ in the 2D {\it (a)} and
3D {\it (b)} cases. Dotted lines (right scale) show the carrier density
$x_{FM}(J)$ in the ferromagnetic regions.}
\label{fig:tensions}
\end{figure}

This conclusion is expected
to affect the
properties of the system in a case where the ferromagnetic phase occupies
a considerable fraction of the sample. The long-range
Coulomb interaction dictates that instead of a single charged macroscopic
droplet of a ferromagnet, a Wigner crystal-like array of smaller droplets is
formed\cite{Debye}. Provided that these droplets are large enough\cite{large}
and well separated in space, the shape of each droplet is governed
by the results reported above.
The small-$J$ regime of  $w-1 \ll 1$, and hence
circular droplets, is predicted for Eu-based
semiconductors\cite{Nagaevbook}.
As more carriers are doped into such a system, the geometry of the
state eventually changes from that of disconnected droplets
of a (metallic) ferromagnet embedded in an  antiferromagnetic insulator
to the inverse situation of a connected metallic phase\cite{Nagaevbook}.


The regime of larger $J\agt 0.015$ is expected to be realized in the CMR
manganates\cite{Jcom}. The corresponding $w$ exceeds $1.25$, leading to the
following consequences:

\noindent
(i) {\it Hysteretic effects in transport}. When the ferromagnetic
droplets are not connected an applied voltage gives rise to internal
electrostatic fields. Connected conduction paths could then form in order
to lower the associated field energy.
When subsequently the voltage is applied along another direction,
these paths should restructure accordingly. Due to the dependence of the
boundary energy on orientation, this cannot
occur smoothly, but necessarily involves discrete transitions (over energy
barriers) and hysteresis.
Experimentally, switching and memory effects
in the resistivity of CMR manganates near the percolation threshold
have been observed both in thin films\cite{films} and in 3D single
crystals\cite{bulk}.
We emphasize that such properties
arise here as intrinsic to the system, as opposed to more conventional
effects of impurity pinning of the boundary.

\noindent
(ii) For a fixed area fraction of the ferromagnet, $x/x_{FM}$, the ratio
$\Lambda=(\Omega_s-\Omega_{dr})\sqrt{\epsilon}/(e^2 W_d)^{1/2}$, where
$\Omega_{dr}$ and $\Omega_s$  are the energies of a droplet phase and of a 
phase composed of diagonal ferro- and antiferromagnetic stripes,
depends solely on the droplet geometry\cite{prb03} (here, $\epsilon$ is the 
dielectric constant). Variational studies\cite{Nagaevbook,prb03} suggest
that for $w=1$ (circular droplets) and $x/x_{FM}\approx 1/2$ the quantity
$\Lambda$ is only slightly larger than 0. While leaving this issue for future 
study, we note that since $\Omega_{dr}\sqrt{\epsilon}/(e^2 W_d)^{1/2}$ is 
expected to be larger for diamond-shaped droplets
(due to a higher value of boundary energy per particle), it is possible that 
for large $w$ a {\it diagonal stripe phase} proves energetically
favorable ($\Lambda < 0$) in a region near $x/x_{FM}=1/2$.
In a broader physical context, droplet shape effects
may complement other mechanisms suggested\cite{Khomskii} to
stabilize a stripe phase in the CMR manganates.

\noindent {\bf The 3D case}. As seen from Fig. \ref{fig:tensions} {\it (b)},
the energies of the vertical and diagonal boundaries follow each other closely
(with $W_d>W_v$ for $J<0.00019$) throughout the relevant range of the
values of $J$. Thus, using the same
arguments as above (cf. Fig. \ref{fig:circlesquare}) we conclude that the
optimal droplet shape remains close to that of an octahedron (formed
by diagonal boundaries)
with its vertices ``shaved off'' by vertical and horizontal planes.
Energetically,
this geometry is nearly
indistinguishable from that of a sphere. In such a case, one can expect
based on
the variational results for spherical droplets \cite{Nagaevbook,Nagaev72}
that an alternating arrangement of ferro- and antiferromagnetic planar slabs
is unstable for all values of $x/x_{FM}$. However,
the boundary energy even in 3D is not at all orientation-independent.
This is illustrated by the dashed-dotted line in
Fig. \ref{fig:tensions} {\it (b)}, which shows the energy $W_{110}$
of a boundary perpendicular to the (1,1,0) direction\cite{110wall},
\begin{equation}
W_{110}= \frac{1}{2\sqrt{2}}\int_{-\pi}^\pi\frac{dk_y}{2\pi}
\int_{-\pi/2}^{\pi/2}\frac{dk_z}{2\pi}\int_{-3}^{\mu}\xi(\epsilon,k_y,k_z)
d\epsilon -\frac{3J}{\sqrt{2}}.
\label{eq:wall110}
\end{equation}
Here, the
function $\xi$ is given by Eq. (\ref{eq:xivert})
with $Q=2 \cos (k_y/2)$ and $\epsilon_\pm=\epsilon \pm \cos k_z$.
At small $J$, we find
\begin{equation}
W_{110}\approx \frac{3^{8/5}5^{4/5}}{16\cdot 2^{2/5}}\pi^{3/5}J^{4/5}
\!-\frac{6-\sqrt{3}}{\sqrt{2}}J.
\end{equation}
As illustrated by Fig. \ref{fig:tensions} {\it (b)},
$W_{110}$ exceeds
both $W_d$ and $W_v$ by some 20\% in an extended range of values of $J$.
Thus, our conclusion on hysteresis due to abrupt switching of the conduction
paths {\it persists in 3D} (at least for some directions of the path).
This has an enhanced significance because
in 3D, point defects are less likely to interfere with the restructuring
of conduction paths.

We note that in a phase-separated system quenched (chemical) disorder
that leads to variations in the on-site energies also
tends to sub-divide large ferromagnetic areas into droplets.
Since the boundary energy originates from an integral of
the carrier phase shift over the entire Fermi sea,
it is expected to be only weakly sensitive to the
detailed structure of individual wavefunctions (as long as the disorder
is not too strong).
Hence, we expect the orientational dependence
of the boundary energy to persist in the presence of
such inhomogeneities (and of many-body effects due to the short-range part of
the Coulomb interaction\cite{limitcom}). This is illustrated by
numerical-simulation results for a 2D phase-separated double-exchange magnet
with disorder \cite{Majumdar}, showing that the prevailing direction of the
boundaries is clearly along the diagonals.

We note that in order to make quantitative comparison with the experimental
results for the manganates one may have to complement Eq. (\ref{eq:Ham}) with
further terms reflecting orbital and lattice effects inherent to these 
compounds. While these may affect the optimal shape or the preferred 
direction of the boundaries in a phase separated system, our overall 
conclusion on the presence of an orientational dependence of the boundary 
energy is expected to persist. As demonstrated above for the simple model 
(\ref{eq:Ham}),
this general conclusion follows from the presumed short-wavelength (abrupt)
nature of the boundary, combined with sufficiently large band-filling.
The latter ensures that the short-wavelength (on the scale of the lattice constant) 
properties of the electron gas are sensitive to the orientation with respect to 
the lattice. Including phonon or orbital degrees of freedom may modify
this short wavelength behavior, but is rather unlikely to eliminate
it altogether.

As we have previously mentioned, experimental observations
of non-ohmic behavior and memory effects in the resistivity of
CMR manganates near the percolation threshold exist both for 3D single
crystals\cite{bulk} and for thin films\cite{films}. While qualitatively this
is in line with our expectations, it would be desirable to have direct imaging
of the spatial structure of the underlying phase-separated state (as provided
by magnetic force microscopy). Such measurements could also be used to
search for any putative stripe phases, especially in 2D.
We are not aware of similar transport studies of the Eu-based magnetic
semiconductors, and indeed we do not expect intrinsic non-linear effects 
to be prominent
in this case. However, this point should also be confirmed by experiments.

We acknowledge helpful discussions with O. Agam, G. Jung,
and V. Markovich, as well as  the support of the ISF (under the Centers
of Excellence Program), the BSF (grant No. 2004162), and  the
Israeli Absorption Ministry.

~

\appendix*
\section{}

\begin{figure}[ht!!!]
\includegraphics[angle=0,width=3.41in]{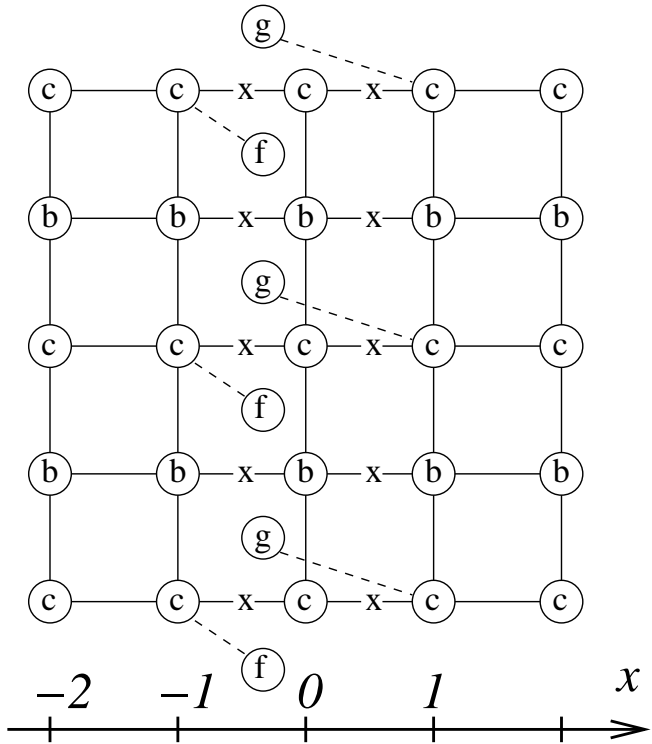}
\caption{The effect of the perturbation, Eq. (\ref{eq:Vlocal}), in the 2D case. Carrier hopping
along the lattice links marked ``x'' is eliminated, and the new bonds (dashed lines) are
established.} \label{fig:cut-connect}
\end{figure}

In the following we briefly outline the calculation of the ``vertical'' boundary energy, Eqs.
(\ref{eq:23Dvert}--\ref{eq:xivert}) using the theory of local perturbations due to I. M. Lifshits
and M. G. Krein\cite{iml}. We first evaluate the double-exchange (electronic kinetic energy)
contribution to the boundary energy. The (auxiliary) unperturbed system consists of two
disconnected components:

\noindent (i) A square (in 3D, cubic) tight-binding network of $N$ sites, which we divide into two
sub-lattices, labeled $b$ and $c$ (see Fig. \ref{fig:cut-connect}), thereby doubling the period in
the direction perpendicular to the $x$ axis (in 3D, this corresponds to a checkerboard arrangement
in the $yz$ plane). Performing a partial Fourier transform in the $y$ (in 3D, $y$ and $z$)
direction, renders the Hamiltonian as
\begin{eqnarray}
{\cal H}_1 &=& \sum_{x, \vec{k}_\bot} \left[
-\frac{1}{2}b^\dagger_{\vec{k}_\bot}(x)b_{\vec{k}_\bot}(x+1)
-\frac{1}{2}c^\dagger_{\vec{k}_\bot}(x)c_{\vec{k}_\bot}(x+1) \right.
\nonumber \\
&&\left.+\epsilon_\bot(\vec{k}_\bot)c^\dagger_{\vec{k}_\bot}(x) b_{\vec{k}_\bot}(x) + {\rm H. c.}
\right],
\end{eqnarray}
where the operators $b$ and $c$ annihilate the two species of spinless fermions on the two
sub-lattices, and $\epsilon_\bot(\vec{k}_\bot)=-\cos k_y (-\cos k_z)$ is the usual tight-binding
dispersion in $D-1$ dimensions.

\noindent (ii) A chain of $\sqrt{N}$ sites (for the 3D case, a plane of $N^{2/3}$ sites),
disconnected from each other (zero hopping amplitudes, hence ${\cal H}_2=0$). We again introduce
two sub-lattices in a checkerboard fashion, and perform the Fourier transform, labeling the two
resulting localized fermion operators $f$ and $g$ with the $D-1$-dimensional momentum index,
$\vec{k}_\bot$.

Next, consider a perturbation of the form
\begin{eqnarray}
&&V=\frac{1}{2} \sum_{\vec{k}_\bot} \left\{ b^\dagger_{\vec{k}_\bot}(0)
[b_{\vec{k}_\bot}(1)+b_{\vec{k}_\bot}(-1)] \right. \nonumber \\
&&+ c^\dagger_{\vec{k}_\bot}(0)[c_{\vec{k}_\bot}(1)+c_{\vec{k}_\bot}(-1)] \nonumber\\
&&\left.-f^\dagger_{\vec{k}_\bot}c_{\vec{k}_\bot}(-1) -g^\dagger_{\vec{k}_\bot}c_{\vec{k}_\bot}(1)
+{\rm H.c.} \right\}. \label{eq:Vlocal}
\end{eqnarray}
The first two terms correspond to disconnecting the $x=0$ chain (plane) from the rest of the
system. An extra site is then connected to every other site at $x=\pm 1$ (last two terms in $V$),
creating a vertical boundary perpendicular to the $x$ axis (see Fig. \ref{fig:cut-connect}). 
Note that hopping is allowed both in the right, and in
the left half-plane (half-space). Therefore, the change in energy of the system
 due to the perturbation 
(when supplemented with the appropriate superexchange contribution) is equal to twice the energy 
of a vertical boundary between an antiferromagnet (where the hopping amplitude
vanishes) and a ferromagnet, 
as pictured in Fig. \ref{fig:walls} {\it(b)}.

Since the transverse momentum $\vec{k}_\bot$ is preserved by $V$, the variables separate. 
For a given $\vec{k}_\bot$, the perturbation is  localized at $x=0,\pm 1 $ and takes the form
\begin{equation}
V_{\vec{k}_\bot}= \sum_{i=1}^{6} A_i a^\dagger_i a_i \; ,
\end{equation}
where $2A_i= \{-\sqrt{3},-1,1,\sqrt{3},-\sqrt{2},\sqrt{2}\}$ and $a_i$ are suitable combinations of
the original fermion operators that satisfy $a_ia^\dagger_j +a^\dagger_j a_i=\delta_{ij}$. The 
spectral shift function, Eq. (\ref{eq:xivert}), is then evaluated as\cite{iml}
\begin{equation}
\xi(k_\bot,\epsilon)=-\frac{1}{\pi} {\rm Arg ~ Det} \left\{\hat{1}-M_{ij}A_j\right\} \, ,
\label{eq:generaliml}
\end{equation}
with
\begin{eqnarray}
M_{ij}&=&\int \frac{dk_x}{2 \pi} \left[ \frac{ \langle  a_i|p^+_{\vec{k}}
\rangle
\langle p^+_{\vec{k}} | a_j \rangle} {\epsilon_+ + \cos k_x- {\rm i} 0}
+\frac{ \langle  a_i|p^-_{\vec{k}}\rangle
\langle p^-_{\vec{k}} | a_j \rangle} {\epsilon_- + \cos k_x- {\rm i} 0} \right]
\nonumber \\
&&+\frac{\langle a_i | f_{\vec{k}_\bot}\rangle \langle f_{\vec{k}_\bot}|a_j
\rangle +\langle a_i | g_{\vec{k}_\bot}\rangle \langle g_{\vec{k}_\bot}|a_j
\rangle}{\epsilon -{\rm i} 0}\,.
\end{eqnarray}
Here, the Dirac bra and ket states are defined by $|f_{\vec{k}_\bot}\rangle
=f^\dagger_{\vec{k}_\bot}|0\rangle$ (with $|0\rangle$, the vacuum state), etc.,
and the operators
$p^{\pm}_{\vec{k}}$ are the Fourier components (with respect to $k_x$) of $(c_{\vec{k}_\bot}(x)\pm
b_{\vec{k}_\bot}(x))/\sqrt{2}$. As before $\epsilon_\pm=\epsilon\mp\epsilon_\bot(\vec{k}_\bot)$.

The choice of the Arg branch in Eq. (\ref{eq:generaliml}) is subject to the requirement that
$\xi(k_\bot,\epsilon) \rightarrow 0$ for a vanishing perturbation strength, {\it e.g.,} if the
Hamiltonian is given by ${\cal H}_1+{\cal H}_2+ \alpha V$ with $\alpha \rightarrow 0$. In the 3D
case, we note the appearance of bound (Tamm) surface electron states at the vertical border. These
have the energy $E=\pm\cosh(k_x) +\epsilon_\bot=0$ with 
$|\epsilon_\bot(\vec{k}_\bot)|>1$. For example the wave
function of an $\epsilon_\bot>1$ state in the $x<0$ half of the system differs from zero only on
the $f$ sites ($x=0$), on the $c$ sites with  even $x$, and on the 
$b$ sites with  odd $x$ (see Fig. \ref{fig:cut-connect}), where it is 
proportional to $\exp[k_xx+{\rm
i}k_yy+{\rm i}k_zz]$. The net spectral shift function, Eq. (\ref{eq:generaliml}),
contains contributions from these boundary states, from the extended states in the two 
half systems and also from the states in the disconnected $x=0$ chain/plane 
(see Fig. \ref{fig:cut-connect}). The latter needs to be subtracted when 
calculating the desired boundary energy. The resulting 
$\xi(\epsilon,\epsilon_\bot)$ is given by Eq. (\ref{eq:xivert}).

By counting the number of antiferromagnetic links and the associated energy change, one arrives at
the final expression, Eq. (\ref{eq:23Dvert}). The energy of a $\{110\}$ boundary, Eq.
(\ref{eq:wall110}), can be evaluated in a similar way.


\begin{thebibliography}{999}
\bibitem{limitcom} In reality, $J_H$ is finite and the strong Hubbard
interaction $U$ must be taken into account.
These facts are not expected to cause a qualitative change in our results,
with the main effect of $U$ being a mean-field renormalization of the model
parameters, cf. Ref. \onlinecite{denis05}.
\bibitem{numericscom} We have verified all our results pertaining to the
2D case via numerical diagonalization of finite clusters.
\bibitem{large} We assume
that the Fermi wavelength in the ferromagnetic phase is much smaller
than the droplet size. In this case, one can omit the effects of
curvature of the boundary, corners and size-dependent
quantum oscillations.
\bibitem{Debye} The size of the droplets should not exceed the
Debye--H\"{u}ckel screening radius (assumed large,
see Ref. \onlinecite{prb03}).
\bibitem{Jcom} The bare $J$, which can be roughly estimated from the N\'{e}el
temperature of fully hole-doped CMR compounds, is enhanced by
inter-subband virtual hopping (cf. Ref. \onlinecite{denis05}).
\end{thebibliography}
\end{document}